\documentclass[aps,prb,reprint,showpacs,superscriptaddress]{revtex4-1}
\usepackage{amssymb,amsmath,amsbsy,graphicx,hyperref,xcolor,csquotes,braket}

\begin{document}
\title{Tunable skewed edges in puckered structures}

\author{Marko M. Gruji\'c}\email{marko.grujic@etf.bg.ac.rs}
\affiliation{School of Electrical Engineering, University of Belgrade, P.O. Box
3554, 11120 Belgrade, Serbia} \affiliation{Department of Physics, University of
Antwerp, Groenenborgerlaan 171, B-2020 Antwerp, Belgium}
\author{Motohiko Ezawa}\email{ezawa@ap.t.u-tokyo.ac.jp}
\affiliation{Department of Applied Physics, University of Tokyo, Bunkyo, Tokyo 113-8656, Japan}
\author{Milan \v{Z}. Tadi\'c}
\affiliation{School of Electrical Engineering, University of Belgrade, P.O. Box
3554, 11120 Belgrade, Serbia}
\author{Fran\c{c}ois M. Peeters}
\affiliation{Department of Physics, University of Antwerp, Groenenborgerlaan
171, B-2020 Antwerp, Belgium}

\begin{abstract}
We propose a new type of edges, arising due to the anisotropy inherent in the
puckered structure of a honeycomb system such as in phosphorene. Skewed-zigzag
and skewed-armchair nanoribbons are semiconducting and metallic, respectively,
in contrast to their normal edge counterparts. Their band structures are
tunable, and a metal-insulator transition is induced by an electric field. We
predict a field-effect transistor based on the edge states in skewed-armchair
nanoribbons, where the edge state is gapped by applying arbitrary small
electric field $E_z$. A topological argument is presented, revealing the
condition for the emergence of such edge states.
\end{abstract}
 \maketitle

\section{Introduction}

Recently black phosphorus attracts much attention since the experimental
demonstration of a field-effect transistor (FET).\cite{li14} Several
experimental and theoretical papers appeared in the last year, which has
demonstrated its properties and potential applications.
\cite{rodin14,liu14,gomez14,xia14,qiao14,koenig14} Graphene is not viable for
post-silicon electronics due to the lack of a band gap, while the band gap of
black phosphorus can be tuned by changing the number of layers or applying an
electric field.\cite{gomez14,rudenko14,kim15} Black phosphorus could
potentially serve a whole range of purposes for FET, while retaining higher
electron mobility than transition-metal dichalcogenides.

The graphene analogue of black phosphorus is phosphorene,
\cite{liu14,rudenko14} a monolayer sheet of phosphorus atoms arranged in a
honeycomb lattice. The structure of phosphorene is puckered, where the
emergence of distinct zigzag ridges leads to highly anisotropic properties.
While the majority of research is focused on bulk phosphorene,
\cite{low14,cakir14,yuan15,pereira15,chaves15,tahir15,srivastava15} it is of interest to
analyze the properties of phosphorene nanoribbons.\cite{peng14,du15,carvalho14} Zigzag
graphene nanoribbons have zero-energy edge states, while armchair ones can be
semiconducting or metallic, depending on their
width.\cite{nakada96,fujita96,ezawa06} Recently, it was shown that a similar
picture holds true for phosphorene nanoribbons, where zigzag nanoribbons host a
quasi-flat band detached from the bulk, corresponding to edge states near the
Fermi level,\cite{ezawa14} while armchair nanoribbons are semiconducting, and
become metallic under a sufficiently strong electric field.
\cite{sisakht15,guo14}

\begin{figure}
\centering
\includegraphics[width=8.6cm]{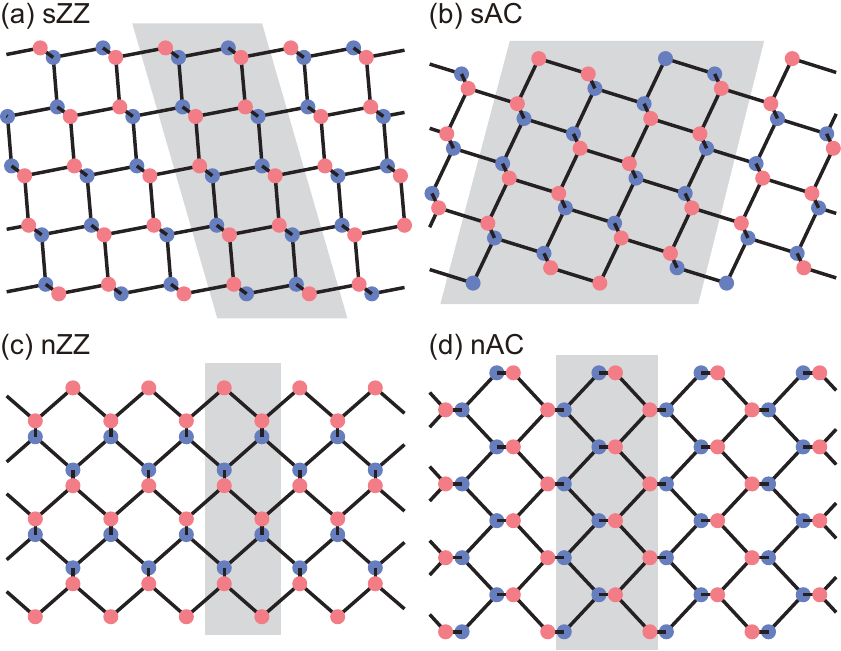}
\caption{Illustrations of
(a) sZZ nanoribbon with $N_{\text{sZZ}}=5$,
(b) sAC nanoribbon with $N_{\text{sAC}}=8$,
(c) nZZ nanoribbon with $N_{\text{nZZ}}=5$ and
(d) nAC nanoribbon with $N_{\text{nAC}}=8$
periodic along the parallel direction. Unit cells are outlined for each case by the gray area. The colors
denote the ridge structure, with red sublayer atoms shifted vertically
above the blue sublayer atoms.}
\label{fig1}
\end{figure}

In this paper, we propose a new type of edges in the anisotropic honeycomb
lattice system without the $C_{3}$ rotational symmetry. Phosphorene is the best
example. On the one hand, the planar honeycomb structure has the $C_3$
rotational symmetry. On the other hand, only the $C_2$ symmetry exists in the
puckered structure due to the anisotropy, which allows us to introduce a new
type of zigzag or armchair edges. In particular, we may cut the honeycomb sheet
so that the zigzag (armchair) direction intersects the puckered ridges under an
angle other than $0^{\circ}$ ($90^{\circ}$), as illustrated in Fig.
\ref{fig1}(a) and (b). We refer to these nanoribbons as skewed-zigzag (sZZ) and
skewed-armchair (sAC) nanoribbons. 
These are a new type of edges.
In contrast we refer to the normal
nanoribbons as normal-zigzag (nZZ) and normal-armchair (nAC) nanoribbons, as
shown in Fig. \ref{fig1}(c) and (d). By calculating the band structure, we find
that the sZZ nanoribbons are semiconducting while the sAC nanoribbons are
metallic. Namely we have found an unexpected duality between a set of the
skewed edges and a set of the normal edges: If the normal nanoribbon is
metallic, the skewed one is insulating, and vice versa. This duality has a
topological origin. We next show that the properties of these skewed edges are
electrically tunable. First, an electric field along the width of the
nanoribbon can close (open) the band gap in sZZ (sAC) nanoribbons. Second, it
is remarkable that an arbitrarily small electric field perpendicular to the
sheet, obtainable by vertical gating, can open a band gap in sAC nanoribbons.
It would become an insulator under $E_z\approx 26$ mV\r{A} at room temperature
for phosphorene.
This is not the case for nZZ nanoribbons, where band gap opening is possible
only for certain nanoribbons subjected to extremely large fields. Although we
use phosphorene as an explicit example, our arguments are applicable to any
materials with anisotropic nearest neighbor hoppings.

\begin{figure}
\centering
\includegraphics[width=8.6cm]{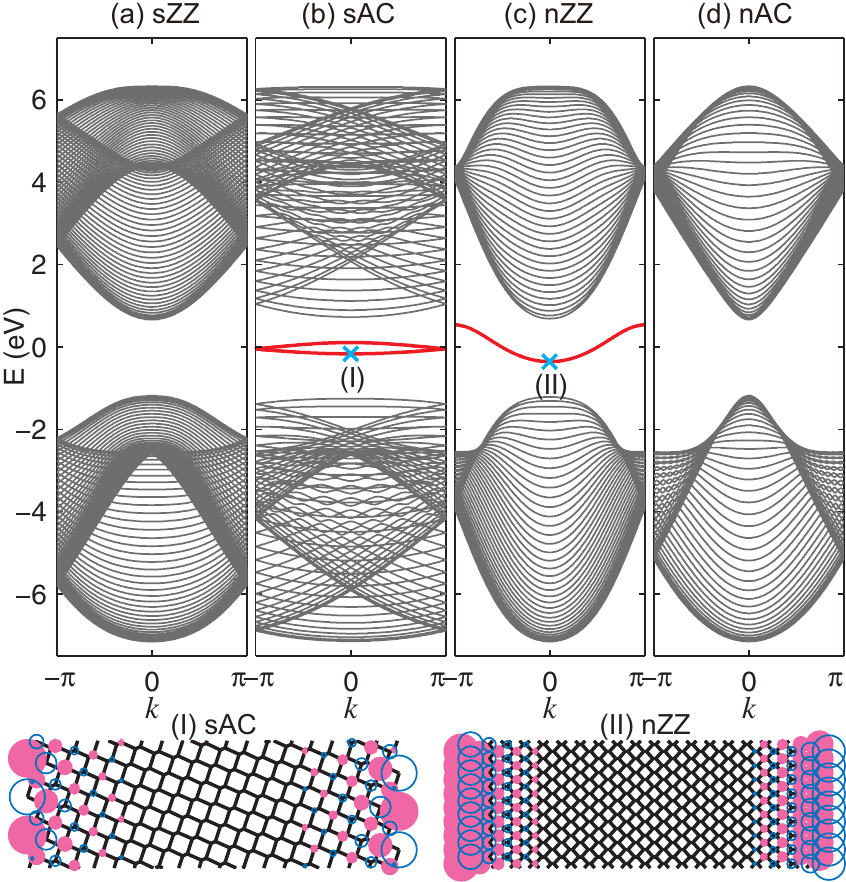}
\caption{Band structure of (a) sZZ, (b) sAC, (c) nZZ and (d) nAC nanoribbons with width $N=40$.
There are QFBs (red curves) in the sAC and nZZ nanoribbons, but not in the sZZ and nAC nanoribbons.
Panels (I) and (II) depict the real space localization of states belonging to QFBs at the crossed points in (b) and (c), respectively.
The magenta disks (blue circles) denote the probability weight on the atoms in the upper (lower) sublayer.
The probability weight between the left and right edges are identical.}
\label{fig2}
\end{figure}

\section{Band Structure}

The tight-binding model is adequate to elucidate the
new physics of the puckered system. It helps us to capture the essential
physics irrespective of material details. The model Hamiltonian reads
\begin{equation}\label{eq:Hamiltonian}
H=\sum_{i\neq j}t_{ij}c^{\dagger}_{i}c_{j}+\sum_i U_i c^{\dagger}_{i}c_{i},
\end{equation}
where $t_{ij}$ are parameters describing the hopping between the lattice sites
$i$ and $j$, while $U_i$ is the on-site energy induced by the electric field. We
show the lattice structure of nanoribbons in Fig. \ref{fig1}. Note that the sZZ
(sAC) direction is not parallel (normal) to the puckered ridges which are shown
by red and blue disks. It is convenient to define the width $N$ of the normal
(skewed) nanoribbon by one half (quarter) of the number of atoms in the unit cell,
which we denote also by $N_{\text{nZZ}}$ or $N_{\text{nAC}}$ ($N_{\text{sZZ}}$
or $N_{\text{sAC}}$) appropriately. For each nanoribbon we perform a Fourier
transform to the momentum basis, diagonalize the corresponding Hamiltonian,
and obtain in the band structure.

To display explicitly the band structure we employ the tight-binding Hamiltonian derived recently by Rudenko et
al. for phosphorene,\cite{rudenko15} which contains $10$ hopping parameters $t_{ij}$ up to the site separation $\sim5.5$ \AA.
It reproduces well the band structure near the Fermi level.

Typical band structures of sZZ, sAC, nZZ and nAC nanoribbons are shown for the
case of width $N=40$ in Fig. \ref{fig2}. On the one hand, no edge states are
present and a large band gap is opened in the sZZ nanoribbon. This is highly
contrasted to the nZZ nanoribbon [Fig. \ref{fig2}(d)], where there are
edge-localized states.\cite{ezawa14} In this sense, the band structure of the
sZZ nanoribbon resembles that of the nAC nanoribbon.\cite{sisakht15} On the
other hand, two quasi-flat bands (QFBs) appear in the middle of the band gap
for the band structure of the sAC nanoribbon, as shown in Fig. \ref{fig2}(b).
It is analogous to the nZZ nanoribbon [Fig. \ref{fig2}(c)]. Indeed, it is
possible to verify topologically that QFBs in the sAC nanoribbon share the same
topological origin as QFBs in the nZZ nanoribbon.

\begin{figure}
	\centering
	\includegraphics[width=8.6cm]{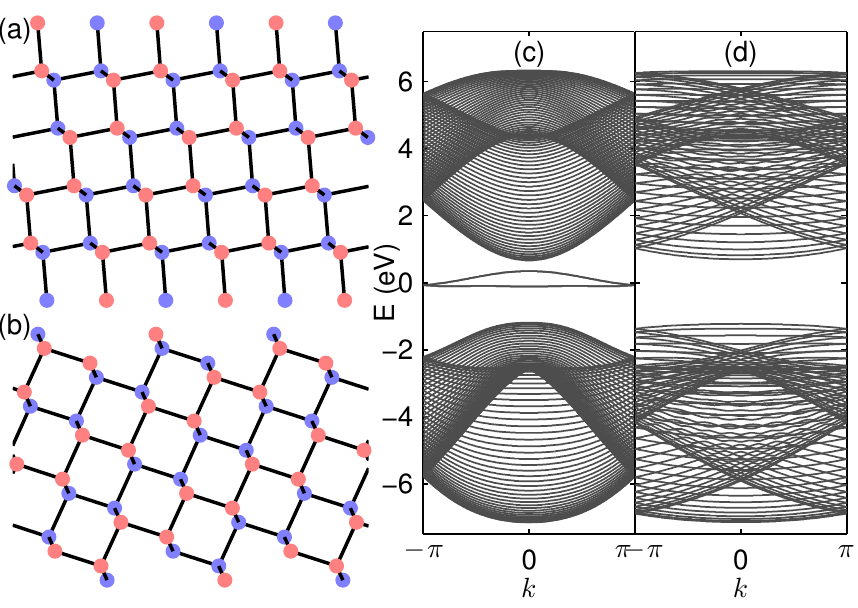}
	\caption{The lattice of (a) skewed-bearded nanoribbons, (b) skewed-dangled nanoribbons while the corresponding band structure is shown in panels (c) and (d), respectively.}
	\label{beard-dangle}
\end{figure}

A comment is in order with respect to another set of edges. Interestingly, adding one neighboring atom to	the outermost atoms along the edge, turns skewed-zigzag (armchair) nanoribbon into a skewed-bearded (dangled) nanoribbon, with (without) quasi-flat bands. Most likely, these nanoribbons are unstable and prone to passivization, but are nevertheless important in providing insight into the anisotropic honeycomb nanoribbon model. In Fig. \ref{beard-dangle}(a) and (b) we show the real space layout of these nanoribbons. The band structures of such nanoribbons are shown in Fig. \ref{beard-dangle}(c) and (d), respectively. Note that they are virtually identical to the band structures of the related nanoribbons they are derived from, apart from the fact that now edge states appear when they were not present before, and vice versa.

\section{Topological Arguments}

We reveal the topological origin of the emergence and the absence of edge states. The essential structure of the anisotropic honeycomb lattice is given by the Hamiltonian,\cite{ezawa14}
\begin{equation}\label{SimpleModel}
H=(\psi_A^{\dagger}(\mathbf{k}),\psi_B^{\dagger}(\mathbf{k}))
\left(
\begin{array}{cc}
0 & f(\mathbf{k}) \\
f^{\ast }(\mathbf{k}) & 0
\end{array}
\right)
\left(
\begin{array}{c}
\psi_A (\mathbf{k})\\
\psi_B (\mathbf{k})
\end{array}
\right)
\end{equation}
with
\begin{equation}
f(\mathbf{k})=t_{a}e^{ia\left( \frac{k_{x}}{2\sqrt{3}}+\frac{k_{y}}{2}\right)
}+t_{b}e^{ia\left( \frac{k_{x}}{2\sqrt{3}}-\frac{k_{y}}{2}\right)
}+t_{c}e^{-ia\frac{k_{x}}{\sqrt{3}}},
\end{equation}
where three independent parameters $t_a,t_b$ and $t_c$ describe the three nearest neighbor hoppings.
It agrees with the Hamiltonian (\ref{eq:Hamiltonian}) up to the nearest neighbor hopping, which is sufficient to capture the underlying physics.

Caution is needed when we analyze a nanoribbon. First, it is necessary to set
$t_{>}\equiv t_b=t_a\not= t_{<}\equiv t_c$ for the normal edge, and  $
t_{>}\equiv t_c=t_a\not= t_{<}\equiv t_b$ for the skewed edge. Second, although
the phase of the function $f(\mathbf{k})$ is arbitrary for the honeycomb sheet,
it is to be fixed so as to make the coefficient of $t_{j}$ real for the
outermost link according to the type of the edge.\cite{ryu,ezawa14}
We thus use 
\begin{equation}
f_a(\mathbf{k})=e^{-ia\left(
\frac{k_{x}}{2\sqrt{3}}+\frac{k_{y}}{2}\right)}f(\mathbf{k})=t_a + \cdots
\end{equation}
for the nZZ and sZZ edges, and 
\begin{equation}
f_c(\mathbf{k})=e^{ia\frac{k_{x}}{\sqrt{3}}}f(\mathbf{k}) = t_c + \cdots 
\end{equation}
for the nAC and sAC edges.

The momentum $k$ along the nanoribbon is a good quantum number.
Hence we consider a one-dimensional insulator indexed by $k$,
which may be topological or not.
If it is topological for all $k$, the edge is gapless due to the bulk-edge correspondence \cite{ryu} for all these one-dimensinal insulators. Consequently we have predicted the emergence of a perfect flat band in the Hamiltonian (\ref{SimpleModel}).
It is deformed into a realistic QFB when hoppings beyond the nearest neighbor ones are included.\cite{ezawa14}

The topological number is defined by the loop integral along the momentum $k_{\perp}$
orthogonal to the nanoribbon \cite{ryu,ezawa14},
\begin{equation}
N_{\text{wind}}(k)=\frac{1}{2\pi i}\int_{-\pi /a}^{\pi /a}dk_{\perp}\,\partial _{k_{\perp}}\log
F_{k}(k_{\perp}),\label{TopNum}
\end{equation}
where $F_{k}(k_{\perp})=f_a(k_x,k_y)$ with $k=k_x$ and $k_{\perp}=k_y$ for the nZZ and sZZ edges, while $F_{k}(k_{\perp})=f_c(k_x,k_y)$ with $k=k_y$ and $k_{\perp}=k_x$ for the nAC and sAC edges.

\begin{figure*}
\begin{center}
	\includegraphics[width=17cm]{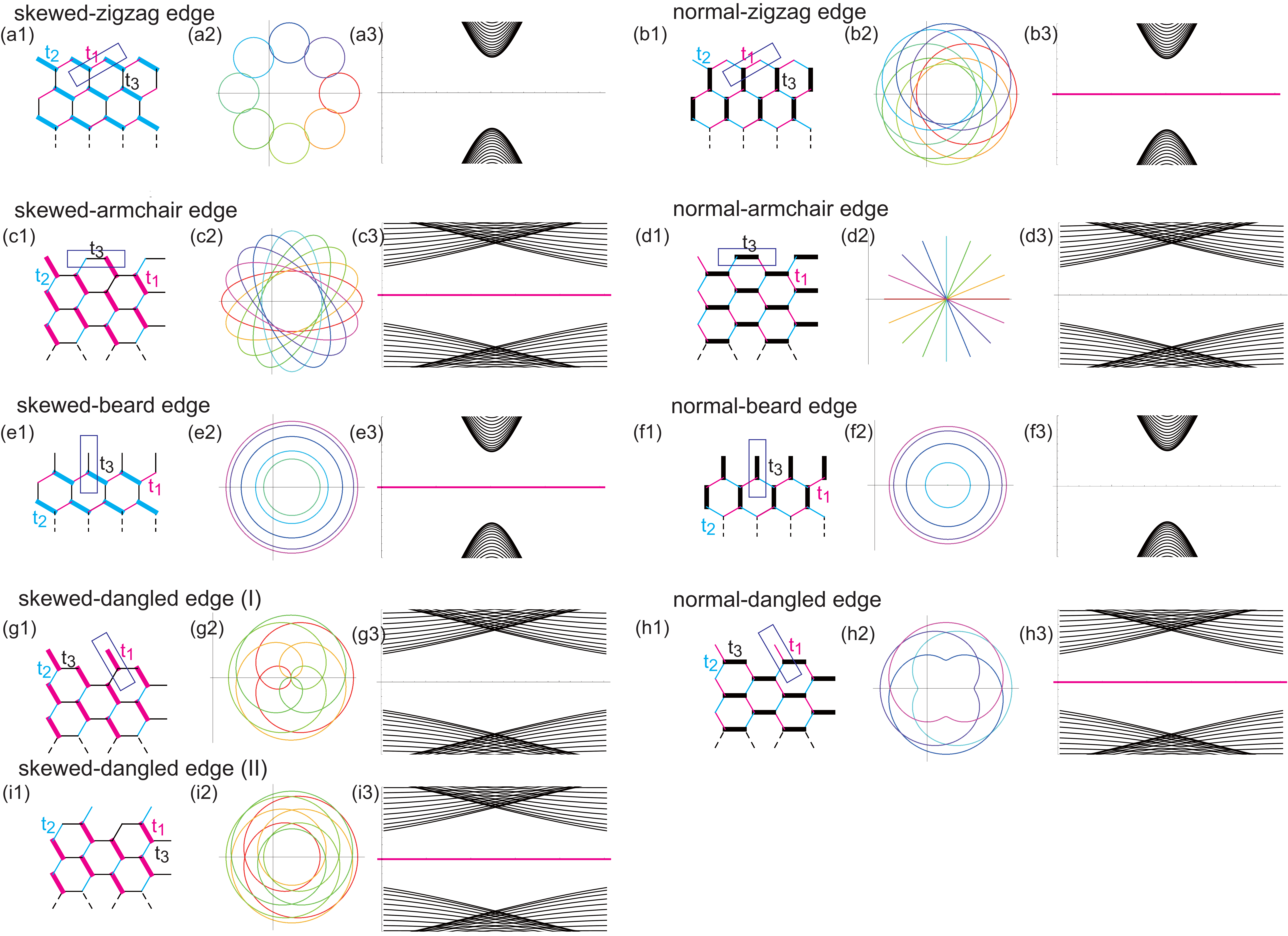}
	\caption{ Illustration of the (a1) skewed-zigzag, (b1) normal-zigzag, (c1)
		skewed-armchair, (d1) normal-armchair edges, (e1) skewed-beard, (f1) normal beard, (g1) skewed-dangled edge (type I), (h1) normal-dangled and (i1) skewed-dangled edge (tyep II). We note that there two inequivalent types of the skewed-dangled edge depending on the type of the dangled atoms. (a2), (b2), (c2), (d2), (e2), (f2), (g2), (h2) and (i2):
		Plots of the loop for the topological number (\protect\ref{TopNum}). The
		horizontal axis is Re[$F_{k}\left( k_{\perp }\right) $], while the
		vertical axis is Im[$F_{k}\left( k_{\perp }\right) $]. The loop encircles
		the origin for all $k$ in the case of the skewed-armchair and normal-zigzag
		edges ($N_{\text{wind}}=1$), while it does not for all $k$ in the case of
		the skewed-zigzag and normal-armchair edges ($N_{\text{wind}}=0$). Color
		indicates the momentum $k$. Red indicates $k=0$ and cyan indicates $k=\protect\pi $.
		(a3), (b3), (c3), (d3), (e3), (f3), (g3), (h3) and (i3): The band structure of nanoribbons
		with the corresponding edges. The edge states are perfectly flat since they
		are calculated based on the simple anisotropic honeycomb model. We have set $t_{>}/t_{<}=2.5$.}
	\label{FigS}
\end{center}
\end{figure*}

The phase should be fixed as follows. For the general zigzag edge
we have
\begin{equation}
F_{k}\left( k_{\perp }\right) =e^{-ia\left( \frac{k_{\perp }}{2\sqrt{3}}
	+\frac{k}{2}\right) }f=t_{a}+t_{b}e^{-iak}+t_{c}e^{-ia\left( \frac{\sqrt{3}k_{\perp }}{2}+\frac{k}{2}\right) },  \label{SMzz}
\end{equation}
while for the general armchair edge we have
\begin{equation}
F_{k}\left( k_{\perp }\right) =e^{ia\frac{k}{\sqrt{3}}}f=t_{a}e^{ia\left(
	\frac{\sqrt{3}k}{2}+\frac{k_{\perp }}{2}\right) }
+t_{b}e^{ia\left( \frac{\sqrt{3}k}{2}-\frac{k_{\perp }}{2}\right) }+t_{c}.  \label{SMac}
\end{equation}
For the general beard edge we have
\begin{equation}
F_{k}\left( k_{\perp}\right) =e^{ia\frac{k_{\perp}}{\sqrt{3}}}f=t_{a}e^{ia\left( 
\frac{\sqrt{3}k_{\perp}}{2}+\frac{k}{2}\right) }+t_{b}e^{ia\left( \frac{\sqrt{3}%
k_{\perp}}{2}-\frac{k}{2}\right) }+t_{c}, \label{SMb}
\end{equation}
while for the general dangled edge we have
\begin{equation}
F_{k}\left( k_{\perp}\right) =e^{-ia\left( \frac{k}{2\sqrt{3}}+\frac{k_{\perp}}{2}%
\right) }f=t_{a}+t_{b}e^{-iak_{\perp}}+t_{c}e^{-ia\left( \frac{\sqrt{3}k}{2}%
-k_{\perp}\right) }. \label{SMd}
\end{equation}
The winding number can be obtained geometrically. The winding number counts
how many times the phase of $F_{k}(k_{\perp })$ surrounds the origin. We find
that the loop is formed in the Re$\left[ F_{k}(k_{\perp })\right] $-Im[$F_{k}(k_{\perp })$]
plane as a function of $k_{\perp }$.
If the loop encircles the origin, $N_{\text{wind}}(k)=1$, while if the loop does not
encircle the origin, $N_{\text{wind}}(k)=0$. The results are shown in
Fig. \ref{FigS}(a2), (b2), (c2) and (d2). Alternatively, we can explicitly evaluate
the winding number using the residual theorem of the complex function, which
are summarized in the following.

\subsection{Zigzag edge}

For the zigzag edge, by inserting (\ref{SMzz}) into (\ref{TopNum}) and setting $z=e^{-ia\frac{\sqrt{3}k_{\perp }}{2}}$, the winding number is rewritten as
\begin{equation}
N_{\text{wind}}(k)=\frac{-i}{2\pi }
\oint dz\frac{t_{c}e^{iak}}{t_{a}+t_{b}e^{-iak}+t_{c}e^{-iak/2}z}.
\end{equation}
A pole exists at
\begin{equation}
z=-\frac{e^{-iak/2}t_{a}+e^{iak/2}t_{b}}{t_{c}}.
\end{equation}
The residual integral gives $N_{\text{wind}}(k)=1$ for $\left\vert
z\right\vert <1$ and $N_{\text{wind}}(k)=0$ for $\left\vert z\right\vert >1$.

Hence, the condition of the emergence of the edge states is explicitly given
by $\left\vert z\right\vert <1$, or
\begin{equation}
t_{a}^{2}+t_{b}^{2}+2t_{a}t_{b}\cos \frac{ak}{2}<t_{c}^{2}
\end{equation}
since
\begin{equation}
z^{2}=\frac{t_{a}^{2}+t_{b}^{2}+2t_{a}t_{b}\cos \frac{ak}{2}}{t_{c}^{2}}.
\end{equation}

For the skewed-zigzag edge, by setting $t_{c}=t_{a}$, this condition is
simplified as
\begin{equation}
t_{b}\left( t_{b}+2t_{a}\cos \frac{ak}{2}\right) <0.  
\end{equation}
It is equals to
\begin{eqnarray}
t_{b} &>&0\text{ and }t_{b}+2t_{a}\cos \frac{ak}{2}<0, \notag\\
t_{b} &<&0\text{ and }t_{b}+2t_{a}\cos \frac{ak}{2}>0.  
\end{eqnarray}
There is no solution for $k$ when $\left\vert z\right\vert <1$,
or $\left\vert t_{b}\right\vert >2\left\vert t_{a}\right\vert $. In conclusion,
edge states do not emerge for $\left\vert t_{b}\right\vert >2\left\vert
t_{a}\right\vert $ in the skewed-zigzag edge as shown in Fig. \ref{FigS}(a3).

For the normal zigzag edge, by setting $t_{b}=t_{a}$, this condition is
simplified as
\begin{equation}
2t_{a}^{2}\left( 1+\cos \frac{ak}{2}\right) <t_{c}^{2}.
\end{equation}
A solution exists for all $k$ when $\left\vert t_{c}\right\vert
>2\left\vert t_{a}\right\vert $. In conclusion, a perfect flat band emerges
for $\left\vert t_{c}\right\vert >2\left\vert t_{a}\right\vert $ in the
normal-zigzag edge as shown in Fig. \ref{FigS}(b3).

\subsection{Armchair edge}

For the armchair edge, by inserting (\ref{SMac}) into (\ref{TopNum}) and setting $z=e^{ia\frac{k_{\perp }}{2}}$, the winding number is rewritten as
\begin{equation}
\begin{split}
N_{\text{wind}}(k)=\frac{-i}{2\pi}
\oint \frac{dz}{z}\frac{t_{a}e^{ia\frac{\sqrt{3}k}{2}}z^{2}
	-t_{b}e^{ia\frac{\sqrt{3}k}{2}}}{t_{a}e^{ia\frac{\sqrt{3}k}{2}}z^{2}
	+t_{b}e^{ia\frac{\sqrt{3}k}{2}}+t_{c}z}.
\end{split}
\end{equation}
Poles exist at
\begin{equation}
z_{0}=0,\quad z_{\pm }=\frac{1}{2t_{a}}e^{-ia\frac{\sqrt{3}}{2}k}
\left(-t_{c}\pm \sqrt{t_{c}^{2}-4e^{ia\sqrt{3}k}t_{a}t_{b}}\right) .
\end{equation}
We find
\begin{equation*}
\text{Res}\left[ z_{0}\right] =-1,\qquad \text{Res}\left[ z_{\pm }\right] =1.
\end{equation*}

For the skewed-armchair edge, by setting $t_{c}=t_{a}$, two poles satisfies
$\left\vert z_{\pm }\right\vert <1$ and $N_{\text{wind}}(k)=1$
for $\left\vert t_{b}\right\vert >2\left\vert t_{a}\right\vert $. In conclusion,
perfect flat bands emerge for $\left\vert t_{b}\right\vert >2\left\vert
t_{a}\right\vert $ in the skewed-armchair edge as shown in Fig. \ref{FigS}(c3).

For the normal armchair edge, by setting $t_{b}=t_{a}$,
we find $N_{\text{wind}}(k)=0$ for $\left\vert t_{c}\right\vert >2\left\vert t_{a}\right\vert $
since $\left\vert z_{+}\right\vert <1$ and $\left\vert z_{-}\right\vert >1$
for $t_{c}>0$ and $\left\vert z_{+}\right\vert >1$ and $\left\vert
z_{-}\right\vert <1$ for $t_{c}<0$. In conclusion, edge states do not
emerge for $\left\vert t_{c}\right\vert >2\left\vert t_{a}\right\vert $ in
the normal armchair edge as shown in Fig. \ref{FigS}(d3).

To summarize, the topological number (\ref{TopNum}) for each type of nanoribbon,
for all $k$ with $|t_>|>2|t_<|$, we find
\begin{equation}
\begin{tabular}{|c|c|c|c|c|c|}
\hline
& zigzag & armchair &beard & dangled (I) &dangled (II)\\ \hline
normal & 1 & 0  &0&1&$\times$\\ \hline
skewed & 0 & 1 &1&0&1\\ \hline
\end{tabular}
\label{table}\ .
\end{equation}
It manifests the duality that the existence and absence of the edge states are opposite in the skewed and normal nanoribbons.
Note that our topological arguments are applicable to any materials with anisotropic nearest neighbor hoppings
with the condition $|t_>|>2|t_<|$ in (\ref{SimpleModel}). Furthermore, we would like to stress that the topological nature of edge states in phosphorene is not the same as edge states in conventional 2D topological insulators. In particular, the edge states in phosphorene are less robust, and more sensitive to disorder.

\begin{figure}
\centering
\includegraphics[width=8.6cm]{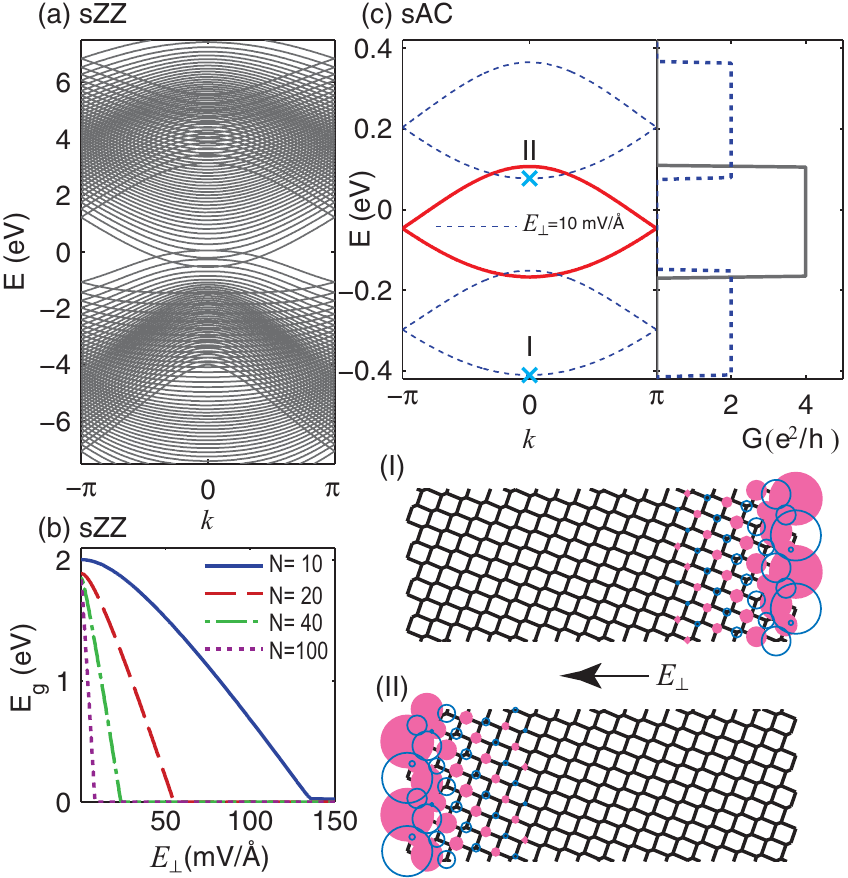}
\caption{(a) Band structure of $N=40$ sZZ nanoribbon for $E_{\perp}=30$ mV/\AA.
(b) The impact of electric field along the sZZ nanoribbon width on the band gap for
$N=10, 20, 40$ and $100$.
(c) Quasi-flat bands of sAC nanoribbons without external field (red curves), and with side gating
$E_{\perp}=10$ mV/\AA (blue dotted curves).
The resulting conductance reveals the degeneracy and the band gap of the bands.
Panels (I) - (II) depict the real space localization
of states belonging to QFBs denoted by two cyan crosses in panel (c).
The magenta disks (blue circles) denote the probability weight on the atoms in the upper (lower) sublayer.
The wave functions are localized at one edge.
}
\label{fig3}
\end{figure}

\begin{figure}
\centering
\includegraphics[width=8.6cm]{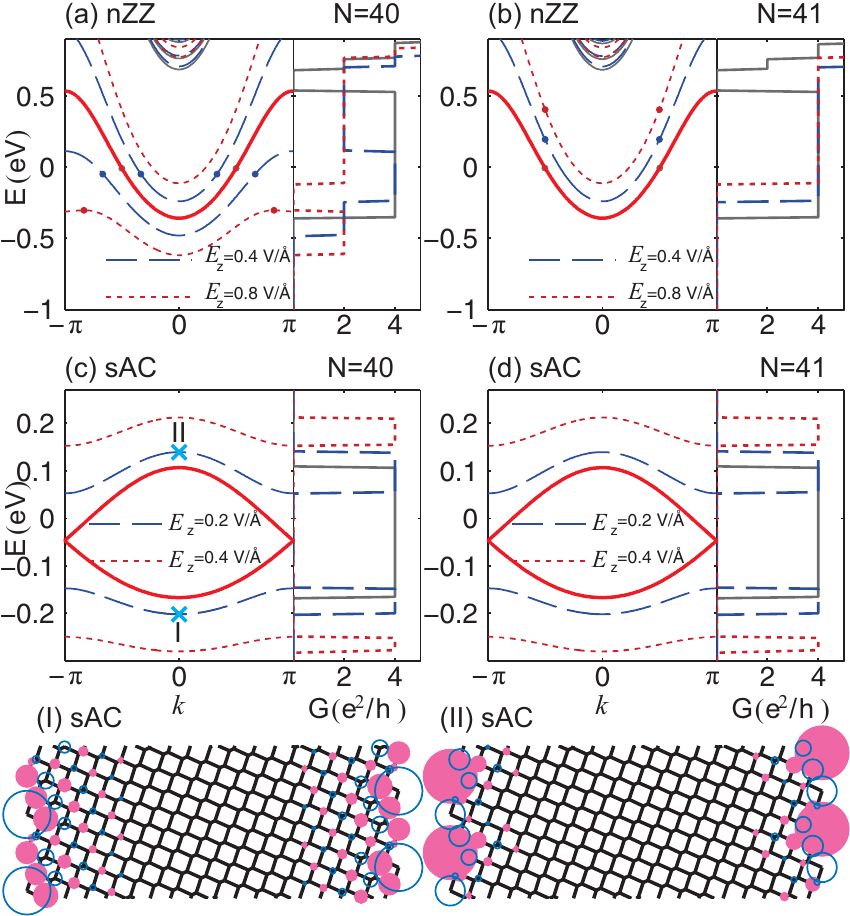}
\caption{Band structures of (a) even ($N=40$) and (b) odd ($N=41$)
nZZ nanoribbons, and (c) even ($N=40$) and (d) odd ($N=41$)
sAC nanoribbons, in the absence (solid red curves), and presence
of perpendicular electric field $E_z$ (dotted curves). The corresponding
 conductance is shown in the right panels for all cases. Note the lack of
parity effect in skewed armchair nanoribbons.
Panels (I) and (II) depict the real space localization of states belonging to QFBs.
The probability weight between the upper and lower sublayers are imbalanced due to the Stark effect induced by $E_z$.}
\label{fig4}
\end{figure}

\section{Lateral Gating}

It is a standard technique to employ an electric field
to control the band gap. We start with a discussion on the effects of lateral
gating. We can close the band gap of sZZ nanoribbons by applying an electric
field $E_{\perp}$ along the nanoribbon width [Fig.\ref{fig3}(a)]. Due to a
large Stark effect, the conduction (valence) bands are shifted downwards
(upwards) in energy, resulting in an insulator-metal transition.

The critical field is smaller for wider nanoribbons due to the following two reasons:
i) wider nanoribbons experience larger potential variation along the width direction due to the field, making them more sensitive to smaller fields;
ii) the band gap of wider nanoribbons in the absence of external fields is
smaller, and asymptotically approaches the bulk band gap for wide enough naroribbons, which is $1.84$ eV for phosphorene.\cite{rudenko15} These two effects
are illustrated in Fig. \ref{fig3}(b) for $N_{\text{sZZ}}=10, 20, 40$ and $100$.

In Fig. \ref{fig3}(c) we take a closer look at the QFBs under an applied
electric field $E_{\perp}$. One can see that sAC nanoribbons can be transformed
into an insulator by lateral gating since the two-fold degenerate edge states
split into opposite directions, thus breaking the double degeneracy. The wave
functions become localized at the nanoribbon edge for the occupied and
unoccupied QFB, as shown in the panels (I) and (II), respectively. The
conductances are switched off, which implies that sAC nanoribbons act as an FET
by $E_{\perp}$.

\section{Vertical Gating}

We proceed to investigate the effects of the electric field perpendicular to the honeycomb sheet ($E_z$).
First we analyze nZZ nanoribbons. In Fig. \ref{fig4} we show the
band structure, and the corresponding conductance of (a) $N_{\text{nZZ}}=40$
and (b) $N_{\text{nZZ}}=41$ nanoribbons, for $E_z=0$ (black full curves), $E_z=0.4$
V/\AA (blue dotted curves) and $E_z=0.8$ V/\AA (red dashed curves).
Intriguingly, we find that the behavior near the Fermi level at half filling
(depicted by a dot) depends crucially on the parity of $N_{\text{nZZ}}$.

When $N_{\text{nZZ}}$ is an even number, the two outermost ridges of the
nanoribbon, which support most of the weight of the edge states, belong to
different sublayers. Therefore, applying $E_z$ will differentiate the two
degenerate QFBs, and split them into opposite directions. For a sufficiently
strong $E_z$ an energy gap appears between the two bands, and since only the
lower band is filled at half-filling, this constitutes an FET action. The
critical field is high ($\sim660$ mV/\AA), which is independent of the width,
unless the nanoribbons are very narrow. The reason for such a high critical
field is the relatively large bandwidth of the QFBs. The edge state at the
momentum $k=\pi$ is almost localized at the outermost ridge, while the wave
function penetrates into the bulk as $k$ reaches $0$. Accordingly, the shift of
the energy at $k=\pi$ is maximum and that at $k=0$ is minimum.

On the other hand, if $N_{\text{nZZ}}$ is an odd number, the outermost ridges
exist in the same plane as shown in Fig.\ref{fig1} (c). Accordingly, both bands
of the edge states shift in the same direction by applying $E_z$, keeping the
degeneracy intact, as can be seen in Fig. \ref{fig4}(b). Hence, a gap cannot be
induced by $E_z$ at half-filing. Namely it is necessary that $N_{\text{nZZ}}$
is even to make a FET. Furthermore, an extremely large electric field is
necessary to open a gap, which is an obstacle for FET operation.

Finally we analyze sAC nanoribbons. In Fig. \ref{fig4}(c) and (d) we depict the
band structure of $N_{\text{sAC}}=40$ and $N_{\text{sAC}}=41$ nanoribbons,
respectively. One can see that $E_z$ opens a band gap, by shifting the
conduction (valence) band up (down). The double degeneracy of the folded bands
are preserved, which is evident from the conductance. The corresponding edge
states with finite $E_z$ become predominantly localized on the upper (lower)
sublayer, as shown in panels (I) and (II) of Fig. \ref{fig4}. Accordingly, the
edge states corresponding to the positive-curvature (electron-like) valence
bands move to lower energy, while those corresponding to the negative-curvature
(hole-like) conduction bands move to higher energy. Hence, the electric field
separates the electron- and hole-like states and bands apart. This is the Stark
effect. Importantly, a gap opens under arbitrarily small electric field. We
find $E_z\approx 26$ mV\r{A} in order to open a gap corresponding to the room
temperature. Moreover, note that there are no parity issues in sAC nanoribbons;
they all respond in the same manner to an external electric field, regardless
of the width. In conclusion, sAC nanoribbons may well be suited for practical
devices.

\section{Discussion and the Conclusion}
In our analysis we have assumed atomically precise edges, 
about which we make a few comments. (i) Structural relaxation and the
effect of dangling bonds can be properly treated by means of first-principles
calculations. Previous first-principles calculations have shown \cite{guo14}
that the quasi-flat bands are robust for pristine normal nanoribbons when
taking into account the structural relaxation and dangling bonds. We might
expect that our results are also valid for pristine nanoribbons. (ii) Edge
states will emerge when there are some regions whose edges are locally precise
although the global edge structure is disordered. Indeed, edge states of
graphene have been observed even when the edge is not perfectly precise.
\cite{tao11} (iii) With respect to the impact of the electric field, the shift
of the QFB will be robust in the presence of disorder since it is due to the
total imbalance of the upper and lower electron distribution, which is
independent of the details of the edge roughness.

We also comment on the numerical values such as the critical electric field we
have derived for phosphorene. Since they are derived based on the tight-binding model, they can
be modified by first-principles calculations. Nevertheless, our results give an order
of magnitude estimate, which will benefit future experimental works.

In conclusion, we have proposed a new type of edges, the sZZ and sAC edges,
that can appear in puckered honeycomb systems, and revealed intriguing
properties inherent to nanoribbons possessing these edges. The sAC nanoribbon
is particularly interesting, since it has QFBs whose origin is topological, and
since a gap opens by applying an arbitrary small electric field $E_z$.
This property may allow a FET to be designed.

\begin{acknowledgments}This work was supported by the Serbian Ministry of Education, Science and Technological Development, and the Flemish Science Foundation (FWO-Vl). M.E. thanks the support by the Grants-in-Aid for Scientific Research from MEXT KAKENHI (Grant~Nos.~25400317~and~15H05854).
\end{acknowledgments}

\end{document}